\documentclass[conference]{IEEEtran}

\usepackage{cite}
\usepackage{amsmath,amssymb,amsfonts}
\usepackage{graphicx}
\usepackage{booktabs}
\usepackage{multirow}
\usepackage{threeparttable}
\usepackage{textcomp}
\usepackage{xcolor}
\usepackage{url}
\usepackage[hidelinks]{hyperref}%
\usepackage[capitalise]{cleveref}
\crefname{figure}{Fig.}{Figs.}
\crefname{table}{Table}{Tables}
\crefname{equation}{Eq.}{Eqs.}
\crefname{section}{Section}{Sections}
\crefname{subsection}{Section}{Sections}
\crefname{definition}{Definition}{Definitions}
\crefname{examplebase}{Example}{Examples}

\newtheorem{definition}{Definition}
\usepackage[most]{tcolorbox}
\tcolorboxenvironment{definition}{enhanced,breakable,colback=white,colframe=black!55,boxrule=0.4pt,arc=1pt,left=3pt,right=3pt,top=1pt,bottom=1pt,before skip=4pt,after skip=4pt}

\newcommand{\rate}[3]{\textcolor{black!55}{#1/#2E$#3$}}
\newcommand{\tcite}[1]{\textcolor{blue!70!black}{\cite{#1}}}
\newcommand{\two}[2]{\begin{tabular}[t]{@{}c@{}}#1\\[-2pt]{\scriptsize #2}\end{tabular}}

\newtheorem{examplebase}{Example}
\newenvironment{example}[1][]{\ifx\relax#1\relax\begin{examplebase}\else\begin{examplebase}[#1]\fi\itshape}{\end{examplebase}}
\newcommand{\ket}[1]{|#1\rangle}

\begin{document}
\bstctlcite{IEEEexample:BSTcontrol}

\title{\vspace{-0.5cm}Generalizing and Combining Check Constructions:\\ Design-Space Exploration for Fault-Tolerant\\ Quantum State Preparation\vspace{-0.2cm}}

\author{\IEEEauthorblockN{Ludwig Schmid\textsuperscript{1} and Robert Wille\textsuperscript{1,2}}
\IEEEauthorblockA{\textsuperscript{1}Chair for Design Automation, Technical University of Munich, Munich, Germany\\
\textsuperscript{2}Munich Quantum Software Company GmbH, Munich, Germany} \vspace{-1.2cm}}

\maketitle

\begin{abstract}
Fault-tolerant quantum computing needs logical states prepared such that single faults cannot spread into uncorrectable errors.
Post-selection achieves this with added checks that detect dangerous faults and discard the run.
How the checks are measured sets the depth and operation count of the circuit, and the best trade-off depends on the hardware.
Existing automated methods, however, each use one check construction and yield fixed trade-offs.

We first introduce \emph{Block Checks}, in which several checks share one fault-tolerantly prepared auxiliary block, which places them and existing constructions in a single design space.
Second, to explore that space, we propose a Pareto tree search that adds checks and auxiliary blocks, merges auxiliaries, and edits the circuit.
On quantum codes of distance up to seven, the search finds best-known circuits for many new operation--depth trade-offs and surpasses existing methods.
The code is open source as part of the Munich Quantum Toolkit (MQT) at \url{https://github.com/munich-quantum-toolkit/qecc}, and all circuits are available in the QECirc library at \url{https://qecirc.com}.
\end{abstract}

\section{Introduction}
Fault-tolerant (FT) quantum computing needs reliably prepared logical states, both for initialization and as auxiliary states throughout the computation.
For small and moderate codes, post-selected preparation~\cite{knillQuantumComputingRealistically2005} is attractive due to its low resource overhead and competitive error rates~\cite{gotoMinimizingResourceOverheads2016,pehamAutomatedSynthesisFaulttolerant2025,zenQuantumCircuitDiscovery2025,forlivesiFlagOriginModular2025,crigerAutomatedFlagbasedFaulttolerant2026,weilandtSynthesisFaulttolerantState2026,poorSpidercssScalableFaulttolerant2026,schmidDeterministicFaulttolerantState2025,pehamOptimizingFaulttolerantCat2026}.
Post-selected preparation is therefore typically among the first FT primitives demonstrated on hardware~\cite{ryanandersonRealizationRealtimeFaulttolerant2021,postlerDemonstrationFaulttolerantUniversal2022,bluvsteinLogicalQuantumProcessor2024,paetznickDemonstrationLogicalQubits2024,perlinFaulttolerantExecutionErrorcorrected2026,postlerDemonstrationFaulttolerantSteane2024}.
Especially for this near-term hardware, the cost of such a circuit matters, since every extra operation introduces additional faults, and every extra layer adds idle errors.

Automated methods for post-selected preparation exist~\cite{pehamAutomatedSynthesisFaulttolerant2025,forlivesiFlagOriginModular2025,weilandtSynthesisFaulttolerantState2026,pehamOptimizingFaulttolerantCat2026,poorSpidercssScalableFaulttolerant2026,crigerAutomatedFlagbasedFaulttolerant2026}, but most commit to a certain kind of check construction.
Some place Bare Checks, with flags where needed, after or within the preparation~\cite{pehamAutomatedSynthesisFaulttolerant2025,crigerAutomatedFlagbasedFaulttolerant2026}.
They require few operations, making them the most widely used approach in experiments~\cite{ryanandersonRealizationRealtimeFaulttolerant2021,postlerDemonstrationFaulttolerantUniversal2022,paetznickDemonstrationLogicalQubits2024,perlinFaulttolerantExecutionErrorcorrected2026}, but their sequential coupling tends to make the circuits deep.
Another uses Steane-Type Checks~\cite{weilandtSynthesisFaulttolerantState2026,postlerDemonstrationFaulttolerantSteane2024,bluvsteinLogicalQuantumProcessor2024}, whose circuits are very shallow but need many additional qubits and operations.

Each of these methods yields specific points in the trade-off between operations and depth.
Which circuit is best, however, depends on the hardware, its parallelism, and its coherence time.
Neutral-atom processors, for example, run many gates in parallel and profit most from low depth~\cite{everedHighfidelityParallelEntangling2023,bluvsteinLogicalQuantumProcessor2024}, but the shallow Steane-Type circuits pay with many operations, which partly cancels their depth advantage.
The best circuit for a platform may thus lie between these points, and no existing method explicitly explores this trade-off.
Combining the constructions is hard, because a shared auxiliary must itself be prepared fault-tolerantly, a nested preparation problem.

We address this gap with a two-step approach:
\begin{itemize}
  \item We introduce \emph{Block Checks} and their grouped variant, shared auxiliary blocks inspired by the syndrome extraction approach from Huang and Brown~\cite{huangBetweenShorSteane2021}, to post-selected FT state preparation.
    Block Checks measure several checks through one fault-tolerantly prepared auxiliary and interpolate between Cat Checks and Steane-Type Checks.
    In state preparation, they allow for shallow circuits with fewer operations than Steane-Type Checks.
  \item This places most existing constructions and Block Checks in a single design space.
    To explore this large design space, we propose a Pareto tree search that combines different check constructions in one circuit and constructs optimized circuits at various depths.
\end{itemize}

We evaluate the search on CSS codes of distance up to seven.
The search fills the space between the existing approaches and, at the depth of every published circuit, finds circuits with fewer operations.
The implementation is open source as part of the Munich Quantum Toolkit (MQT)~\cite{willeMQTHandbookSummary2024}, and all constructed circuits are available at QECirc.com~\cite{schmidQECircCommunitydrivenFAIR2026}.

\section{Check Constructions}\label{sec:background}

\begin{figure*}[t]
  \centering
  \vspace{-0.5cm}
  \includegraphics[width=0.96\textwidth]{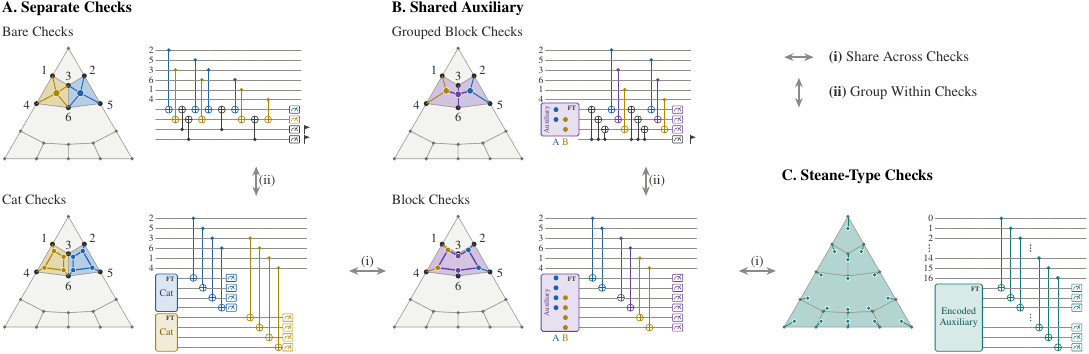}
  \caption{Ways to measure checks, for the $[[17,1,5]]$ color code.
  The columns correspond to \Crefrange{sec:separate}{sec:steane}.
  Each panel shows where the auxiliary system (colored) couples to the code, and the circuit with only the coupled data qubits drawn.
  Boxes are fault-tolerantly prepared auxiliary states.
  The panels are connected by the two merges of \Cref{sec:shared} and their inverses: (i)~Share Across Checks (between columns) and (ii)~Group Within Checks (between rows).}
  \label{fig:checks}
\end{figure*}

This section reviews how post-selected preparation detects faults and the existing check constructions, and \Cref{sec:shared} introduces Block Checks and Grouped Block Checks to state preparation, with the $[[17,1,5]]$ color code as an example.

A CSS code $[[n,k,d]]$ has X- and Z-type stabilizers~\cite{calderbankGoodQuantumErrorcorrecting1996,steaneMultipleparticleInterferenceQuantum1996}, and its logical basis states are stabilizer states.
A circuit of resets and CNOTs can prepare such a state, but it generally will not be fault tolerant.
A single fault can propagate through the CNOTs to a high-weight error, which could be dangerous to the code.
In the post-selected scheme, measured checks are added to the circuit, and any detection discards the run, which results in a repeat-until-success protocol.

\begin{definition}[Fault tolerance]\label{def:ft}
Let $C$ prepare a state of an $[[n,k,d]]$ CSS code, let $\mathcal{S}$ be the stabilizer group of the prepared state (code stabilizers and the logicals that fix it, e.g., $Z_L$ for $\ket{0}_L$), let $t=\lfloor (d-1)/2\rfloor$, and let \mbox{$\mathrm{wt}_{\mathcal{S}}(E)=\min_{S\in\mathcal{S}}\mathrm{wt}(ES)$} be the weight of an error $E$ modulo $\mathcal{S}$.
A set $F$ of faults in $C$ with resulting errors $E_X$ and $E_Z$ is \emph{dangerous} if $\mathrm{wt}_{\mathcal{S}}(E_X)>|F|$ or $\mathrm{wt}_{\mathcal{S}}(E_Z)>|F|$.
Following~\cite{pehamAutomatedSynthesisFaulttolerant2025}, $C$ is \emph{fault tolerant} if every dangerous set $F$ with $|F|\le t$ is detected.
\end{definition}

\begin{example}
The $[[17,1,5]]$ color code of \Cref{fig:checks} has eight X- and eight Z-type stabilizers on 17 qubits and distance 5, so $t=2$ by \Cref{def:ft}.
Throughout the following, we prepare the logical $\ket{0}$ state and measure the two overlapping Z-type checks $A$ on qubits $\{2,3,5,6\}$ and $B$ on $\{1,3,4,6\}$.
\end{example}

To measure a check, data qubits are coupled to auxiliary qubits, which are then measured.
\Cref{fig:checks} shows the constructions discussed in the following, which achieve fault tolerance at different operation and depth costs.

\subsection{Separate Checks}
\label{sec:separate}
The simplest construction uses one auxiliary qubit per check, coupled to the check's whole support one CNOT at a time~\cite{pehamAutomatedSynthesisFaulttolerant2025}, as in \Cref{fig:checks}A (top).
We refer to this structure as \textbf{Bare Checks}.
A fault on such an auxiliary can spread to several data qubits, a typical hook error, which can be detected by flags~\cite{chaoQuantumErrorCorrection2018,chamberlandFlagFaulttolerantError2018}.
Bare Checks can be placed after the preparation, as in layered verification~\cite{pehamAutomatedSynthesisFaulttolerant2025}, or inside it~\cite{crigerAutomatedFlagbasedFaulttolerant2026}, where they detect errors before these spread.
The ILP selection of Criger et al.~\cite{crigerAutomatedFlagbasedFaulttolerant2026} combines both by choosing a low-cost set of deterministic gauge operators.

\textbf{Cat Checks} in the style of Shor~\cite{shorFaulttolerantQuantumComputation1996} are the other classical way.
Each check gets a cat state that couples transversally to the data.
In general, the cat state needs a fault-tolerant preparation, as discussed in \Cref{sec:hooks}, which can be achieved via optimized constructions~\cite{prabhuFaulttolerantSyndromeExtraction2023,pehamOptimizingFaulttolerantCat2026,khesinSpidercatOptimalFaulttolerant2026}.
Bare Checks need few operations but grow deep with the check weights, while Cat Checks are shallow but need more qubits and CNOTs.
Both have in common that each check is measured by a separate auxiliary system.

\begin{example}
In \Cref{fig:checks}A (top), each check uses one auxiliary qubit and one flag, and the 4 interaction CNOTs of each check run one after another.
With Cat Checks (bottom), $A$ and $B$ each use a 4-qubit cat state, prepared fault-tolerantly (FT): 8 interaction CNOTs and 8 auxiliary qubits plus those of the cat preparations are required, but with pre-built cat states the interaction has only depth 2, since qubits 3 and 6 are shared.
\end{example}

\subsection{Shared Auxiliary}
\label{sec:shared}
This separation, however, is not a requirement.
Huang and Brown~\cite{huangBetweenShorSteane2021} showed for syndrome extraction that auxiliary qubits can be merged in two ways, shown in \Cref{fig:checks}:
\begin{enumerate}
  \item[(i)] \textbf{Share Across Checks}: merge auxiliary qubits that touch the same data qubit.
    Fewer interaction CNOTs are needed, but the auxiliary becomes larger and more entangled.
  \item[(ii)] \textbf{Group Within Checks}: merge auxiliary qubits that count toward the same checks.
    As a result, the auxiliary qubit touches several data qubits.
    This makes the interaction more complicated while giving a smaller auxiliary.
\end{enumerate}
Starting from Bare Checks, ungrouping (the inverse of (ii)) expands each auxiliary to the full check size and yields the known Cat Checks of \Cref{sec:separate}.
Applying (i) to Cat Checks yields what we refer to as \textbf{Block Checks}: several checks share one auxiliary block, and each data qubit couples to it once transversally.
Applying (ii) to a block then yields what we call \textbf{Grouped Block Checks}: the auxiliary shrinks, but the coupling becomes more complicated and spans several layers.
Cat Checks and Steane-Type Checks are the two extremes of Block Checks, over one check and over all checks of one type~\cite{gormanTransversalSyndromeExtraction2026}.
Block Checks over several but not all checks, and Grouped Block Checks, have not been used for automated post-selected state preparation so far, and we introduce both here.
Grouped Block Checks require extra care~\cite{huangBetweenShorSteane2021}, as both the auxiliary state \emph{and} the interaction must be protected.
\Cref{sec:nested} treats each block preparation as a nested design, protected like any other state.

\subsection{Steane-Type Checks}
\label{sec:steane}
Applying merge (i) to all checks of one type (either X or Z) gives \textbf{Steane-Type Checks}~\cite{steaneActiveStabilizationQuantum1997}: an encoded auxiliary copies the errors of the data through one transversal CNOT layer and thereby measures all stabilizers of that type at once.
Such encoded auxiliaries are even harder to prepare fault-tolerantly, as preparing them is the original preparation problem again.
Paetznick and Reichardt~\cite{paetznickFaulttolerantAncillaPreparation2012} for the Golay code and Weilandt et al.~\cite{weilandtSynthesisFaulttolerantState2026} for general CSS codes use three additional carefully crafted auxiliary states that detect each other's errors and the data's errors.
Steane-Type Checks are very shallow, but preparing multiple full states and their transversal coupling require many CNOTs~\cite{weilandtSynthesisFaulttolerantState2026}.

\begin{example}
In \Cref{fig:checks}C, one fault-tolerant 17-qubit auxiliary measures all eight Z-type checks with 17 CNOTs in one layer.
\end{example}

\subsection{Auxiliary Faults}
\label{sec:hooks}
What makes fault-tolerant state preparation hard, however, is not only the choice of a suitable set of checks, but also that the auxiliary systems themselves can be faulty.
A fault on the auxiliary can break the check in two ways, and the auxiliary must itself be guarded against both:
\begin{enumerate}
  \item \textbf{Masking errors}: the fault flips check outcomes and hides a dangerous data error~\cite{weilandtSynthesisFaulttolerantState2026}.
  For a Z-type check, these are $X$ errors on the auxiliary.
  \item \textbf{Hook errors}: the fault spreads back to the data qubits through the CNOTs of the check itself~\cite{chaoQuantumErrorCorrection2018}.
  For a Z-type check, these are $Z$ errors on the auxiliary.
\end{enumerate}
\begin{figure}[t]
  \centering
  \includegraphics{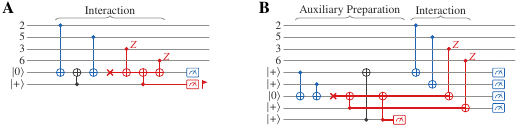}
  \caption{The two kinds of hook error, for check $A$.}
  \label{fig:hooks}
\end{figure}

On a Bare Check, a single fault can flip the outcome and thus hide a data error.
On more complex auxiliaries, a single fault can spread within the auxiliary preparation and flip a correlated pattern of outcomes that may cancel the syndrome of a data error during the transversal coupling~\cite{weilandtSynthesisFaulttolerantState2026}.
Avoiding such masking errors requires careful engineering of the auxiliary system~\cite{weilandtSynthesisFaulttolerantState2026}, or additional or repeated checks.

Hook errors come in two kinds, depending on where the spreading happens.
In an \emph{Interaction Hook}, shown in \Cref{fig:hooks}A for a Bare Check, the auxiliary couples to the data one CNOT at a time, and a fault on it reaches every data qubit coupled afterward.
This way, a single fault can become a high-weight error on the data qubits.
Additional \emph{flags}~\cite{chaoQuantumErrorCorrection2018,chamberlandFlagFaulttolerantError2018} detect it, as in \Cref{fig:checks}A.

Hook errors also occur with a transversal interaction, as in Cat Checks, Block Checks, and Steane-Type Checks (bottom row of \Cref{fig:checks}).
There, each auxiliary qubit couples to only one data qubit, but a fault spreads inside the auxiliary preparation and is then copied to the data.
Such \emph{Auxiliary-System Hooks}, shown in \Cref{fig:hooks}B for a Cat Check, result from the auxiliary preparation instead of the interaction.

Grouped Block Checks lie in between and can have both kinds: a fault on a grouped auxiliary qubit spreads to the data qubits of its group during the coupling, and faults in the smaller auxiliary spread before it.
In general, they need checks on the auxiliary and flags on the coupling, like the flag in \Cref{fig:checks}B (top).

Larger shared auxiliaries are generally harder to prepare against masking and hook errors due to their size and entanglement, so the savings from sharing (i) can be lost to a higher fault-tolerance overhead.

\begin{example}
In \Cref{fig:hooks}A, a $Z$ fault on the auxiliary of $A$ after its coupling to qubit~5 spreads to qubits~3 and~6, and flips the flag.
In \Cref{fig:hooks}B, a $Z$ fault in the 4-qubit cat preparation spreads to the auxiliary qubits coupled to qubits~3 and~6.
In both cases, a single fault on the auxiliary becomes a weight-two error on the data: in A through the interaction, in B through the auxiliary preparation.
\end{example}

\section{Problem and Design Space}\label{sec:problem}
Building on the constructions of \Cref{sec:background}, we first state which circuits we seek and then combine the constructions into a single design space.

\begin{definition}[FT state preparation synthesis]\label{def:problem}
Given a CSS code $[[n,k,d]]$ and a logical state, find circuits of resets, CNOTs and measurements that prepare the state on $n$ data qubits and are fault tolerant by \Cref{def:ft} under post-selection.
\end{definition}

Such circuits differ in their operations (CNOTs and measurements), their depth (the CNOT depth of their schedule), and their acceptance rate.
Operation count and depth generally compete, and the best balance between them depends on the hardware.
We therefore seek a \emph{front}: for each depth $D$, the FT circuits of depth at most $D$ with the fewest operations.

\subsection{Designs}\label{sec:nested}

\begin{figure*}[t]
  \centering
  \vspace{-0.6cm}
  \begin{minipage}[t]{0.805\textwidth}
    \centering
    \includegraphics[height=4.0cm]{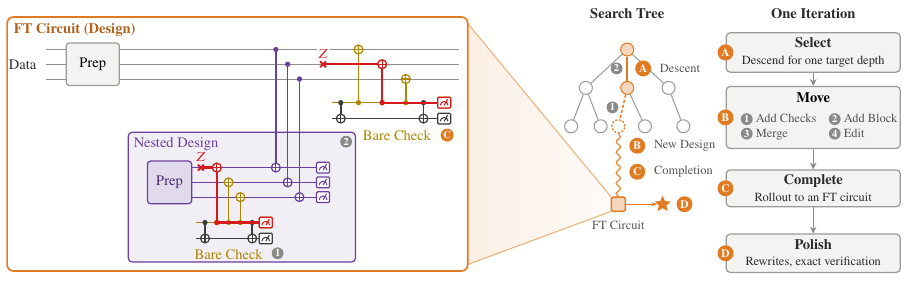}
    \caption{The search and its designs.
    Right: the search tree and one iteration of the search where a node is selected (A), then manipulated with moves (B) and completed to a fault-tolerant design (C) with a final polish (D).
    Left: one such circuit, the design with a nested design of \Cref{ex:nested}, with $Z$ faults and their paths in red.
    Gray numbers: the move families on the path that built it, Add Block (2), then Add Checks (1) inside the nested design, as in \Cref{ex:moves}.
    C: the Bare Check added by the completion.}
    \label{fig:search}
  \end{minipage}\hfill
  \begin{minipage}[t]{0.17\textwidth}
    \centering
    \includegraphics[height=4.0cm]{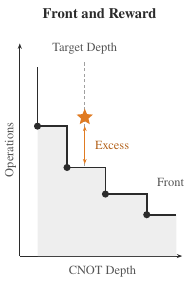}
    \caption{Reward of a new circuit (star) for a certain target depth (dashed).}
    \label{fig:front}
  \end{minipage}
\end{figure*}

We build up a representation of circuits, a \emph{design}, from three observations of \Cref{sec:background}: the best circuit may mix constructions at different positions, merges link the constructions, and a shared auxiliary must itself be prepared fault-tolerantly.

\subsubsection{Placement}
A suitable representation must hold any construction at any position of the data preparation circuit.
It therefore records the construction and position of each check, and further constructions can be added at any position.
The color-code preparation with the Bare Checks $A$ and $B$ placed after it, as in \Cref{fig:checks}A (top), is such a circuit.

\subsubsection{Merges}
Since the constructions are linked by the merges (i) and (ii), applying a merge to a design must again yield a design.
Ungrouping $A$ and $B$ gives the Cat Checks of \Cref{fig:checks}A (bottom), and sharing these the Block Check of \Cref{fig:checks}B (bottom).

\subsubsection{Nesting}
The Block Check, however, needs its 6-qubit auxiliary prepared fault-tolerantly against the masking and hook errors of \Cref{sec:hooks}.
This is \Cref{def:problem} again for the auxiliary state, so the representation is recursive.

\begin{definition}[Design]\label{def:design}
A \emph{design} $\mathcal{D}$ is a preparation circuit with check constructions placed into it.
If a check construction needs a multi-qubit auxiliary state, that state is prepared by a design of its own, a \emph{nested} design.
\end{definition}

\subsubsection{Fault Tolerance}
A design is fault tolerant if its circuit, including all nested designs, satisfies \Cref{def:ft}.
Otherwise, we call it \emph{partial}, since its checks may detect only some of the dangerous faults.
For each construction, this means that its checks detect the dangerous faults of the state it is placed into, and that its auxiliary neither masks these faults nor sends back hook errors that no check or flag detects.

\begin{example}\label{ex:nested}
In \Cref{fig:search} (left), the Bare Check on the data detects a $Z$ fault on the data (red), which the Z-type Block Check cannot see, so no dangerous fault is left.
\end{example}

\subsection{Prior Constructions in This Design Space}\label{sec:prior}
Most existing methods produce designs of a particular form.
Layered verification~\cite{pehamAutomatedSynthesisFaulttolerant2025} and the hand-optimized Steane-code circuit of Goto~\cite{gotoMinimizingResourceOverheads2016} place Bare Checks, with flags where needed, after the preparation.
The ILP selection of Criger et al.~\cite{crigerAutomatedFlagbasedFaulttolerant2026} places them anywhere in it.
Cat Checks~\cite{shorFaulttolerantQuantumComputation1996,pehamOptimizingFaulttolerantCat2026,khesinSpidercatOptimalFaulttolerant2026} are Block Checks over a single check, and Steane-Type Checks~\cite{steaneActiveStabilizationQuantum1997,steaneFastFaulttolerantFiltering2002,reichardtImprovedAncillaPreparation2004,crossComparativeCodeStudy2009,paetznickFaulttolerantAncillaPreparation2012,weilandtSynthesisFaulttolerantState2026} are Block Checks over all checks of one type.

Some existing approaches lie outside this design space.
Flag at Origin~\cite{forlivesiFlagOriginModular2025} places flags on the preparation itself rather than on checks.
SpiderCSS~\cite{poorSpidercssScalableFaulttolerant2026} derives circuits in the ZX calculus and scales to large codes.
Zen et al.~\cite{zenQuantumCircuitDiscovery2025} and Lahiri and Marquardt~\cite{lahiriAgenticAIDiscovering2026} search over whole circuits by reinforcement learning and LLM-driven program evolution rather than over specific check constructions.
Note that the latter uses a different, simpler definition of fault tolerance.

A single design space, on the other hand, lets one explore all these constructions and their combinations.

\section{Search Method}\label{sec:search}

We search the design space of \Cref{sec:problem} by a Pareto tree search~\cite{wangMultiobjectiveMontecarloTree2012}, a multi-objective variant of Monte Carlo tree search~\cite{kocsisBanditBasedMontecarlo2006,browneSurveyMonteCarlo2012} that keeps a front of circuits instead of one best design.

The nodes of the tree are designs as in \Cref{def:design}.
The root is the data preparation without checks, and each child changes one part of its parent by a \emph{move}.
Most nodes are partial designs, which cannot be compared by their costs yet.
The search therefore completes each new design to fault-tolerant circuits and scores these against the front, so a design's score estimates how good the circuits it leads to are.
These scores decide which designs receive the next moves: the search concentrates on subtrees that keep improving the front.

One iteration has the four steps shown in \Cref{fig:search} (right):
\begin{enumerate}
  \item[(A)] \textbf{Select} descends to a design for one target depth.
  \item[(B)] \textbf{Move} creates a new child of it.
  \item[(C)] \textbf{Complete} turns the child into fault-tolerant circuits.
  \item[(D)] \textbf{Polish} rewrites these circuits to optimize them further.
\end{enumerate}

\subsection{Select}\label{sec:select}
The search rates each design by the circuits it has led to and chooses between good and rarely tried designs.
Each descent aims at one target depth $T$, cycling through the depths of the front.

\paragraph*{Rating a circuit}
A new fault-tolerant circuit is rated by how many more operations it needs than the best front circuit of depth at most $T$, as illustrated in \Cref{fig:front}.
For a design $\mathcal{D}$, this is the \emph{relative excess} $e_T = (\mathrm{ops}(\mathcal{D}) - o^*_T)/o^*_T$, where $o^*_T$ is the operation count of that front circuit.
If $\mathrm{depth}(\mathcal{D})>T$, then $r_T(\mathcal{D})=0$, and otherwise the reward is
\begin{equation}\label{eq:reward}
  r_T(\mathcal{D}) =
  \begin{cases}
    1 & \text{if } e_T \le 0,\\
    \max\big(0,\, 1 - e_T/\sigma,\, \tau(1 - e_T/\sigma')\big) & \text{else}.
  \end{cases}
\end{equation}
A circuit on or below the front gets reward 1.
The main term falls to 0 at $\sigma=5\,\%$ more operations, and a small tail with $\tau=0.2$ and $\sigma'=0.2$ keeps circuits up to 20\,\% above the front ranked, since some structures look poor until they are completed and polished.

\paragraph*{Rating a design}
The value $v_T$ of a node is the mean reward of the visits through it, where a visit counts the best reward among the circuits it produced.
Repeated circuits earn nothing, so a subtree that only repeats them loses value.
The reward is added to the node and all its ancestors.

\paragraph*{Choosing the next design}
Each descent picks, at every node $p$, the child $c$ with the highest $v_T(c) + C\sqrt{\ln (N(p)+1) / (N(c)+1)}$ (UCB~\cite{kocsisBanditBasedMontecarlo2006}), which favors good children but also rarely visited ones.
Here, $N$ counts visits, and $C=0.5$.
A node opens a new child only after enough visits, at most $\lceil 2\sqrt{N}\rceil$ children with progressive widening~\cite{couetouxContinuousUpperConfidence2011}, so the search first refines good designs and gradually tries new moves.

\subsection{Move}\label{sec:moves}
A move changes one part of an existing design.
The four move families follow the check constructions of \Cref{sec:background} and the steps that build up a design in \Cref{sec:nested}.
\begin{enumerate}
  \item \textbf{Add Checks}: adds Bare Checks with flags on the dangerous faults, chosen by one ILP round as in~\cite{crigerAutomatedFlagbasedFaulttolerant2026}.
  \item \textbf{Add Block}: adds one block over a set of checks, or over all stabilizers of one type.
    Its auxiliary comes from a cached library of fault-tolerant preparations found earlier or stays unprotected.
    Later moves or the completion then protect it.
    A variant replaces a plain auxiliary by a cancellation-guided preparation after Weilandt et al.~\cite{weilandtSynthesisFaulttolerantState2026}, which avoids masking errors.
  \item \textbf{Merge}: shares, unshares, groups or ungroups the auxiliaries of one check construction, or of two joined ones.
    It either takes one step along merge (i) or (ii) in \Cref{fig:checks} or jumps directly to one of its panels.
  \item \textbf{Edit}: removes one or two parts (a check, a flag, or a block with its auxiliary), shifts a Bare Check to another position in the preparation, or changes the coupling order of a check construction.
\end{enumerate}
Partial designs get all four families, front designs only Merge and Edit.

\begin{example}[Moves]\label{ex:moves}
The circuit of \Cref{fig:search} (left) arises from the root by two moves, tagged in the figure.
Add Block places a block over Z-type checks after the preparation, with an unprotected auxiliary.
Add Checks on that auxiliary's nested design adds a Bare Check with a flag, which protects it.
The completion, described next in \Cref{sec:complete}, then adds the Bare Check with a flag on the data.
\end{example}

\begin{table*}[t]
  \centering
  \vspace{-0.5cm}
  \begin{threeparttable}
    \caption{Best circuits of depth at most $D$ for the logical all-zero state $\ket{0\cdots0}_L$: selected points of our fronts and published prior circuits, as \textbf{ops/CNOTs/auxiliaries} and \textbf{acceptance rate / logical error rate} (any logical qubit) at $p=10^{-3}$.}
  \label{tab:results}
  \scriptsize
  \setlength{\tabcolsep}{2.5pt}
  \renewcommand{\arraystretch}{1.15}
  \begin{tabular*}{\textwidth}{@{\extracolsep{\fill}}r cc c r cc c r cc c r cc@{}}
    \toprule
    \multicolumn{3}{c}{$[[9,1,3]]$ Surface} & & \multicolumn{3}{c}{$[[17,1,5]]$ Color} & & \multicolumn{3}{c}{$[[24,4,5]]$} & & \multicolumn{3}{c}{$[[25,1,5]]$ Surface} \\
    \cmidrule(r){1-3}\cmidrule(lr){5-7}\cmidrule(lr){9-11}\cmidrule(l){13-15}
    $D$ & Ours & Prior & & $D$ & Ours & Prior & & $D$ & Ours & Prior & & $D$ & Ours & Prior \\
    \midrule
    4 & \two{\textbf{12}/\textbf{11}/\textbf{1}}{\rate{0.99}{1.2}{-5}} & \two{\textit{14/12/2}\,\tcite{poorSpidercssScalableFaulttolerant2026}}{\rate{0.99}{1.0}{-5}} &  & 5 & \two{\textbf{100}/\textbf{74}/\textbf{26}}{\rate{0.90}{3.0}{-7}} &  &  & 7 & \two{\textbf{174}/\textbf{134}/\textbf{38}}{\rate{0.83}{5.4}{-7}} &  &  & 5 & \two{\textbf{115}/\textbf{85}/\textbf{30}}{\rate{0.89}{1.4}{-7}} &  \\
    6 & \two{\textbf{12}/\textbf{11}/\textbf{1}}{\rate{0.99}{1.2}{-5}} & \two{\textit{\textbf{12}/\textbf{11}/\textbf{1}}\,\tcite{pehamAutomatedSynthesisFaulttolerant2025}}{\rate{0.99}{7.8}{-6}} &  & 6 & \two{\textbf{88}/\textbf{68}/\textbf{20}}{\rate{0.91}{3.1}{-7}} &  &  & 10 & \two{\textbf{165}/\textbf{131}/\textbf{30}}{\rate{0.84}{5.7}{-7}} &  &  & 6 & \two{\textbf{99}/\textbf{78}/\textbf{20}}{\rate{0.90}{8.7}{-8}} &  \\
     &  & \two{\textit{35/26/8}\,\tcite{forlivesiFlagOriginModular2025}}{\rate{0.97}{1.6}{-5}} &  & 7 & \two{\textbf{77}/\textbf{63}/\textbf{14}}{\rate{0.92}{1.4}{-7}} & \two{\textit{194/143/51}\,\tcite{weilandtSynthesisFaulttolerantState2026}}{\rate{0.80}{9.2}{-8}} &  & 12 & \two{\textbf{156}/\textbf{125}/\textbf{27}}{\rate{0.85}{4.6}{-7}} &  &  & 7 & \two{\textbf{91}/\textbf{73}/\textbf{16}}{\rate{0.91}{1.1}{-7}} & \two{\textit{262/187/63}\,\tcite{weilandtSynthesisFaulttolerantState2026}}{\rate{0.74}{5.0}{-8}} \\
 & & &  & 8 & \two{\textbf{61}/\textbf{53}/\textbf{8}}{\rate{0.94}{1.4}{-7}} &  &  & 14 & \two{\textbf{148}/\textbf{119}/\textbf{22}}{\rate{0.85}{3.3}{-7}} &  &  & 8 & \two{\textbf{82}/\textbf{68}/\textbf{11}}{\rate{0.92}{6.6}{-8}} & \two{\textit{99/78/21}\,\tcite{poorSpidercssScalableFaulttolerant2026}}{\rate{0.92}{2.8}{-7}} \\
 & & &  & 13 & \two{\textbf{55}/\textbf{48}/\textbf{6}}{\rate{0.94}{1.8}{-7}} & \two{\textit{62/53/9}\,\tcite{crigerAutomatedFlagbasedFaulttolerant2026}}{\rate{0.93}{1.4}{-7}} &  & 16 & \two{\textbf{125}/\textbf{109}/\textbf{22}}{\rate{0.87}{2.3}{-7}} &  &  & 9 & \two{\textbf{78}/\textbf{66}/\textbf{12}}{\rate{0.93}{6.8}{-8}} &  \\
 & & &  & 17 & \two{\textbf{55}/\textbf{48}/\textbf{5}}{\rate{0.94}{1.8}{-7}} & \two{\textit{95/74/21}\,\tcite{forlivesiFlagOriginModular2025}}{\rate{0.92}{5.2}{-7}} &  & 17 & \two{\textbf{124}/\textbf{108}/\textbf{14}}{\rate{0.87}{2.3}{-7}} &  &  & 13 & \two{\textbf{74}/\textbf{64}/\textbf{8}}{\rate{0.93}{6.5}{-8}} &  \\
 & & &  & 21 & \two{\textbf{55}/\textbf{48}/\textbf{3}}{\rate{0.94}{1.8}{-7}} & \two{\textit{84/71/8}\,\tcite{pehamAutomatedSynthesisFaulttolerant2025}}{\rate{0.91}{1.6}{-7}} &  & 20 & \two{\textbf{123}/\textbf{107}/22}{\rate{0.87}{2.7}{-7}} & \two{\textit{147/128/\textbf{18}}\,\tcite{crigerAutomatedFlagbasedFaulttolerant2026}}{\rate{0.85}{2.0}{-7}} &  & 20 & \two{\textbf{74}/\textbf{64}/\textbf{8}}{\rate{0.93}{6.5}{-8}} & \two{\textit{120/92/28}\,\tcite{forlivesiFlagOriginModular2025}}{\rate{0.90}{3.4}{-7}} \\
    \midrule
    \multicolumn{3}{c}{$[[20,2,6]]$} & & \multicolumn{3}{c}{$[[23,1,7]]$ Golay} & & \multicolumn{3}{c}{$[[31,1,7]]$ Color} & & \multicolumn{3}{c}{$[[49,1,7]]$ Surface} \\
    \cmidrule(r){1-3}\cmidrule(lr){5-7}\cmidrule(lr){9-11}\cmidrule(l){13-15}
 6 & \two{\textbf{174}/\textbf{130}/\textbf{44}}{\rate{0.82}{4.2}{-7}} &  &  & 8 & \two{\textbf{273}/\textbf{209}/\textbf{62}}{\rate{0.74}{1.3}{-9}} &  &  & 8 & \two{\textbf{282}/\textbf{211}/\textbf{66}}{\rate{0.73}{2.0}{-9}} &  &  & 8 & \two{\textbf{371}/\textbf{276}/\textbf{84}}{\rate{0.67}{1.3}{-9}} &  \\
    7 & \two{\textbf{150}/\textbf{115}/\textbf{33}}{\rate{0.85}{3.9}{-7}} &  &  & 9 & \two{\textbf{270}/\textbf{210}/\textbf{55}}{\rate{0.74}{1.3}{-9}} & \two{\textit{366/297/69}\,\tcite{paetznickFaulttolerantAncillaPreparation2012}}{\rate{0.67}{7.7}{-10}} &  & 10 & \two{\textbf{257}/\textbf{198}/\textbf{56}}{\rate{0.75}{1.9}{-9}} &  &  & 9 & \two{\textbf{345}/\textbf{262}/\textbf{71}}{\rate{0.68}{9.4}{-10}} &  \\
    9 & \two{\textbf{132}/\textbf{105}/\textbf{26}}{\rate{0.86}{5.2}{-7}} & \two{\textit{275/215/58}\,\tcite{weilandtSynthesisFaulttolerantState2026}}{\rate{0.74}{1.8}{-7}} &  & 13 & \two{\textbf{239}/\textbf{190}/\textbf{43}}{\rate{0.77}{1.1}{-9}} &  &  & 11 & \two{\textbf{255}/\textbf{197}/\textbf{54}}{\rate{0.75}{1.7}{-9}} & \two{\textit{402/309/80}\,\tcite{weilandtSynthesisFaulttolerantState2026}}{\rate{0.64}{1.1}{-9}} &  & 14 & \two{\textbf{282}/\textbf{226}/\textbf{44}}{\rate{0.74}{6.1}{-10}} & \two{\textit{295/\textbf{226}/59}\,\tcite{poorSpidercssScalableFaulttolerant2026}}{\rate{0.76}{1.6}{-8}} \\
    12 & \two{\textbf{125}/\textbf{102}/\textbf{21}}{\rate{0.87}{3.5}{-7}} &  &  & 18 & \two{\textbf{231}/\textbf{184}/\textbf{37}}{\rate{0.77}{1.0}{-9}} & \two{\textit{317/237/67}\,\tcite{poorSpidercssScalableFaulttolerant2026}}{\rate{0.73}{4.7}{-8}} &  & 13 & \two{\textbf{248}/\textbf{194}/\textbf{44}}{\rate{0.76}{1.5}{-9}} & \two{\textit{280/211/61}\,\tcite{poorSpidercssScalableFaulttolerant2026}}{\rate{0.76}{1.9}{-8}} &  & 18 & \two{\textbf{274}/\textbf{222}/\textbf{40}}{\rate{0.74}{6.7}{-10}} & \two{\textit{295/226/47}\,\tcite{poorSpidercssScalableFaulttolerant2026}}{\rate{0.76}{1.1}{-8}} \\
    16 & \two{\textbf{118}/\textbf{98}/\textbf{20}}{\rate{0.88}{3.4}{-7}} & \two{\textit{192/145/37}\,\tcite{poorSpidercssScalableFaulttolerant2026}}{\rate{0.83}{1.6}{-6}} &  & 22 & \two{\textbf{223}/\textbf{182}/\textbf{30}}{\rate{0.78}{1.5}{-9}} & \two{\textit{317/237/56}\,\tcite{poorSpidercssScalableFaulttolerant2026}}{\rate{0.73}{4.8}{-8}} &  & 19 & \two{\textbf{226}/\textbf{182}/\textbf{31}}{\rate{0.78}{1.3}{-9}} & \two{\textit{280/211/47}\,\tcite{poorSpidercssScalableFaulttolerant2026}}{\rate{0.76}{1.9}{-8}} &  & 23 & \two{\textbf{270}/\textbf{220}/\textbf{38}}{\rate{0.75}{5.5}{-10}} &  \\
    20 & \two{\textbf{115}/\textbf{98}/\textbf{11}}{\rate{0.88}{2.9}{-7}} &  &  & 28 & \two{\textbf{215}/\textbf{177}/33}{\rate{0.79}{1.0}{-9}} & \two{\textit{255/215/\textbf{32}}\,\tcite{crigerAutomatedFlagbasedFaulttolerant2026}}{\rate{0.76}{5.4}{-10}} &  & & & &  & 42 & \two{\textbf{262}/\textbf{216}/\textbf{36}}{\rate{0.75}{1.1}{-9}} & \two{\textit{295/226/39}\,\tcite{poorSpidercssScalableFaulttolerant2026}}{\rate{0.76}{1.2}{-8}} \\
    26 & \two{\textbf{107}/\textbf{93}/\textbf{10}}{\rate{0.89}{2.4}{-7}} & \two{\textit{166/136/20}\,\tcite{crigerAutomatedFlagbasedFaulttolerant2026}}{\rate{0.83}{1.3}{-7}} &  & 29 & \two{\textbf{215}/\textbf{177}/\textbf{32}}{\rate{0.79}{1.0}{-9}} & \two{\textit{317/237/67}\,\tcite{forlivesiFlagOriginModular2025}}{\rate{0.73}{5.4}{-8}} &  & & & &  & & & \\
    \bottomrule
  \end{tabular*}
  \end{threeparttable}
\end{table*}

\subsection{Complete}\label{sec:complete}
The completion takes the role of the rollout in Monte Carlo tree search and turns a partial design into fault-tolerant circuits.
It tries three ways to deal with the unprotected auxiliaries and records all fault-tolerant results.
Depth weights $\lambda$ in $\mathrm{cost}_\lambda(\mathcal{D}) = \mathrm{ops}(\mathcal{D}) + \lambda\,\mathrm{depth}(\mathcal{D})$ decide which pair is finished:
\begin{enumerate}
  \item \textbf{Protect}: a greedy ILP-based step adds Bare Checks with flags, first on each nested design, then on the data, until no dangerous fault is left.
  \item \textbf{Catch on the data}: the auxiliary stays unprotected, and checks on the data after the blocks detect both the errors the auxiliary sends back and the data errors it masks.
  This can be cheaper than protecting the auxiliary itself.
  \item \textbf{Pair}: the block gets a block of the other type beside it.
  The second block detects the errors the first sends back, and a nested block protects its auxiliary.
  The best pair per depth weight completes the design, and all pairs also become children of the node.
  Pairs typically give shallow completions and so rate designs for low depths.
\end{enumerate}
The completion runs for the smallest, the main and the largest depth weight of a run, by default $\lambda = 0$, $4$ and $20$.
One partial design thus yields circuits at several depths that serve several target depths of the descent.

\subsection{Polish}\label{sec:polish}
The polish optimizes completed circuits by two kinds of rewrites, each kept only if exact verification confirms fault tolerance.
Every completion gets the structured rewrites, and the cheapest completions and all circuits near the front also get the circuit rewrites.

\paragraph*{Structured rewrites} on the design:
(a)~redundancy removal deletes checks, flags or whole blocks that later additions made unnecessary;
(b)~shared flags let one flag qubit serve two Bare Checks, saving a qubit and a measurement;
(c)~flag takeover~\cite{forlivesiFlagOriginModular2025} lets a flag, which holds a copy of its check's auxiliary between its two CNOTs, drive part of the interaction CNOTs, so the check couples on two qubits at once;
(d)~coupling order reorders interaction CNOTs so that overlapping checks run in parallel.

\paragraph*{Circuit rewrites} on the compiled circuit:
(a)~rescheduling reorders gates for lower depth;
(b)~auxiliary removal deletes measured auxiliaries one at a time;
(c)~Bell reduction~\cite{jeonAutomatedReductionFaulttolerant2026}, a fault-equivalent rewrite, removes one qubit of a Bell pair of auxiliaries and saves a qubit, a CNOT and a measurement.

\section{Evaluation}\label{sec:evaluation}
We evaluate the search on CSS codes of distance $d\le 7$ and compare its fronts with published circuits of other methods.

\subsection{Setup}\label{sec:setup}
For each code, we run the search several times, starting from different data preparation circuits, and report the best circuits over all runs.
The runs took 85 core-hours for the $[[9,1,3]]$ code, 1{,}200--1{,}850 per code at distances 5 and 6 and 1{,}650--2{,}150 per code at distance 7.
As baselines, we take all published circuits for these codes and count all CNOTs they need, including transversal ones.
Acceptance and logical error rates are estimated with Monte Carlo sampling using Stim~\cite{gidneyStimFastStabilizer2021} under circuit-level depolarizing noise with $p=10^{-3}$ on preparations, CNOTs, and measurements, and memory noise $p/100$ per idle qubit and CNOT, following~\cite{forlivesiFlagOriginModular2025,poorSpidercssScalableFaulttolerant2026}.

\subsection{Results}\label{sec:results}
At the depth of every published circuit in \Cref{tab:results}, the search finds a circuit of at most that depth with fewer operations, except for one tie for the $[[9,1,3]]$ code at depth~6.
The largest savings are against Steane-Type Checks~\cite{weilandtSynthesisFaulttolerantState2026}.
At depth~7, the $[[25,1,5]]$ surface code needs 91 instead of 262 operations and 16 instead of 63 auxiliary qubits, and the $[[17,1,5]]$ color code needs 77 instead of 194 operations.
The fronts even reach lower depths than any published circuit: depth~5 instead of~7 for the $[[17,1,5]]$ and $[[25,1,5]]$ codes, 6 instead of~9 for $[[20,2,6]]$, 7 instead of~20 for $[[24,4,5]]$, and 8 instead of~9, 11 and~14 for the Golay, $[[31,1,7]]$ and $[[49,1,7]]$ codes.
With fewer operations, the acceptance rates in \Cref{tab:results} match or exceed those of the published circuits, except for the $[[49,1,7]]$ code, where SpiderCSS accepts up to two percentage points more.
The logical error rates are lower than for SpiderCSS~\cite{poorSpidercssScalableFaulttolerant2026} and Flag at Origin~\cite{forlivesiFlagOriginModular2025}, except for the $[[9,1,3]]$ code at depth~4, and about ten times lower at distance~7, but up to three times higher than for the Steane-Type~\cite{paetznickFaulttolerantAncillaPreparation2012,weilandtSynthesisFaulttolerantState2026}, ILP~\cite{crigerAutomatedFlagbasedFaulttolerant2026} and layered-verification~\cite{pehamAutomatedSynthesisFaulttolerant2025} circuits.
This is likely due to undetected sets of $t+1$ faults that are not optimized for in our search but are often detected by the extra checks of these circuits.
Both the acceptance rate and the logical error rate depend heavily on the physical error rate and the exact error model, and are provided here for reference.
Finding a suitable circuit for specific hardware requires careful hardware-specific simulations.

\begin{figure}[t]
  \centering
  \includegraphics[width=\columnwidth]{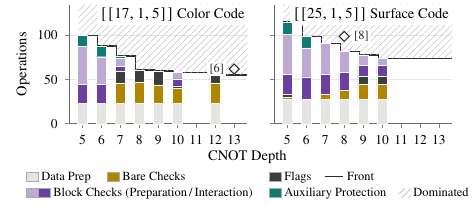}
  \caption{Front circuits by construction.
  Each bar is one front point, its height the operations, split into the data preparation, Bare Checks, flags, Block Checks (auxiliary preparation light, interaction CNOTs dark) and auxiliary protection, the nested designs of the blocks, as read from each circuit.
  Diamonds: published circuits in the shown range, with all others in \Cref{tab:results}.}
  \label{fig:fronts}
\end{figure}

\subsection{Discussion}\label{sec:discussion}
\Cref{fig:fronts} shows which constructions form the front circuits, using the distance-5 color and surface codes as examples.
At the shallow end, most operations outside the data preparation belong to Block Checks and to the nested designs that protect their auxiliaries.
Toward larger depths, the color-code circuits switch to Bare Checks with flags, which need fewer operations but couple one CNOT at a time, while the surface-code circuits keep one block over several checks.
Block Checks thus make the shallow end of the front possible, and the mix of constructions changes gradually along it.

The search and its completion are expensive and do not scale well beyond distance~7 or~9, where post-selected preparation becomes less attractive with falling acceptance rates~\cite{poorSpidercssScalableFaulttolerant2026}.

\section{Conclusion}\label{sec:conclusion}
We introduced Block Checks and their grouped variant to post-selected fault-tolerant state preparation and explored the resulting design space, which contains most existing constructions, with a Pareto tree search.
On CSS codes of distance up to seven, the search finds circuits with fewer operations than every published circuit at its depth and reaches lower depths than any of them.
The code is open source as part of the Munich Quantum Toolkit (MQT)~\cite{willeMQTHandbookSummary2024} at \url{https://github.com/munich-quantum-toolkit/qecc}, and all constructed circuits are available in QECirc~\cite{schmidQECircCommunitydrivenFAIR2026}.

As future work, counting the undetected sets of $t+1$ faults exactly and using the count as an objective alongside or instead of the operation count would allow the search to trade operations for a lower logical error rate.
Also, one could simulate the noise of specific hardware directly to find highly optimized circuits.
Finally, the search could be combined with LLM-driven program evolution~\cite{lahiriAgenticAIDiscovering2026}, for example to propose moves.

\section*{Content Generated by AI}
{\footnotesize
We used Claude Opus 5.5 (Anthropic) as an assistant that supported the implementation of the search and evaluation code, rendered the figures from the authors' sketches, and helped polish the wording.
The method, the experiments, and their interpretation are the authors' own, and the authors reviewed all generated content and take full responsibility for it.\par}

\section*{Acknowledgments}
This work received funding from the European Research Council (ERC) under the European Union's Horizon 2020 research and innovation programme (grant agreement No.~101001318), and from the Deutsche Forschungsgemeinschaft (DFG, German Research Foundation) through the project ``Towards Software for Fault Tolerant Quantum Computing'' (project No.~563402549).
It was also conducted as part of the Munich Quantum Valley, which is supported by the Bavarian state government with funds from the Hightech Agenda Bayern Plus.

\bibliographystyle{IEEEtran}
\bibliography{refs}

\end{document}